\documentclass[aps,prl,twocolumn,showpacs,amsmath]{revtex4}
\usepackage{graphicx}

\begin{document}

\title{ Direct Observation of Hydrogen
Adsorption Sites and Nano-Cage Formation 
in Metal-Organic Frameworks (MOF)}

\author{T. Yildirim and M. R. Hartman}

\affiliation{NIST Center for Neutron Research, National
Institute of Standards and Technology, Gaithersburg, MD 20899}

\begin{abstract}
The hydrogen adsorption sites in MOF5 were determined
using neutron powder diffraction along with first-principles calculations.  
The  metal-oxide cluster is primarily responsible for the adsorption while the
organic linker plays only a secondary role.  Equally important,  at 
low temperatures and high-concentration, H$_{2}$
molecules form unique interlinked high-symmetry nano-clusters with 
intermolecular distances as small as 3.0 \AA \; and 
H$_{2}$-uptake as high as 10-wt\%.  These results hold the key to optimizing
MOF materials for hydrogen storage applications and also suggest
that MOFs can  be used as templates to create artificial
interlinked hydrogen nano-cages with novel properties.
 
\end{abstract}

\date{July 8 2005}
\maketitle

The success of future hydrogen and fuel-cell technologies
is critically dependent upon the discovery of new materials that can store
large amounts of hydrogen at ambient conditions\cite{science-review,zuttel}.
Metal-organic framework (MOF) compounds, which consist of metal-oxide clusters 
connected by organic linkers, are a relatively new class of nano-porous material
that show promise for hydrogen storage applications because of their tunable pore size and 
functionality\cite{mof_design,mof_review,MOF_neutron,mof_hysteretic,mof_expo,mof_calculation,mof_oxy}.
Yet despite numerous experimental studies of hydrogen adsorption in MOF materials,
the nature of the MOF-hydrogen interaction and the manner in which hydrogen molecules 
are adsorbed onto the structure are still unknown.  Answers to these questions hold the 
key to optimizing these materials for practical hydrogen storage applications.  

Here using the difference-Fourier analysis of neutron powder diffraction data 
along with first-principles total-energy calculations, we directly determined the H$_{2}$ 
adsorption sites in MOF5 (the most widely studied  MOF material, which consists 
of ZnO$_4$ clusters linked by 1,4-benzenedicarboxylate (BDC)).  Surprisingly,
the MOF5 host lattice has the enough space available to
hold many hydrogen molecules, up to ~10 wt-\% at low temperatures. 
The ZnO$_4$-cluster is responsible for most of the adsorption 
while the organic linker plays only a secondary role.

\begin{figure}
\includegraphics[scale=0.33]{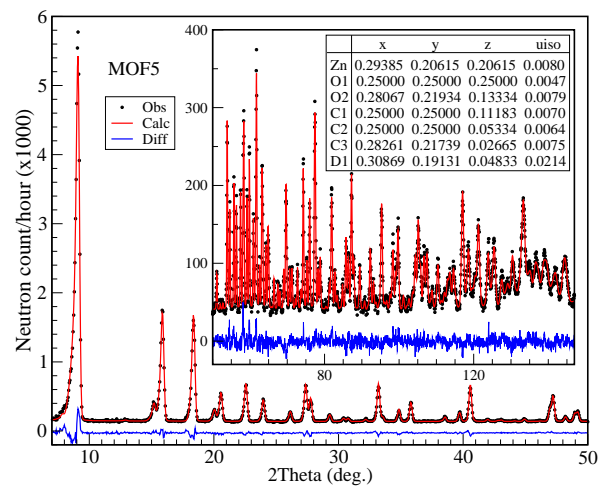}
\caption{
The neutron powder diffraction pattern ($\lambda$=2.08 \AA) 
of the deuterated-MOF5
host lattice at 3.5 K (dots) plus the Rietveld refinement (solid),
using space group Fm${\bar 3}$m and a=25.909 \AA,   
and difference plot (bottom).
The inset shows the high-angle portion of the data along 
with the fractional atomic positions and thermal factors. The
refinement was characterized by $\chi^2$=1.195. }
\label{fig1}
\end{figure}

Equally important, we find that at high-concentration loadings hydrogen 
molecules form unique 3-dimensional (3D) networks of H$_2$ nano-clusters with 
intermolecular distances of 3.0 \AA, which is significantly shorter
than the intermolecular distances  of 3.6 \AA \; in pure 
solid hydrogen\cite{solidh2,H2_pressure}. These findings suggest 
that MOF materials can also be used as templates to create artificial,
interlinked hydrogen nano-cages. Such materials could exhibit 
very unexpected  properties due to confinement effects and 
small intermolecular separation, such as metallic behavior\cite{H2_metal}.

\begin{figure}
\includegraphics[width=80mm]{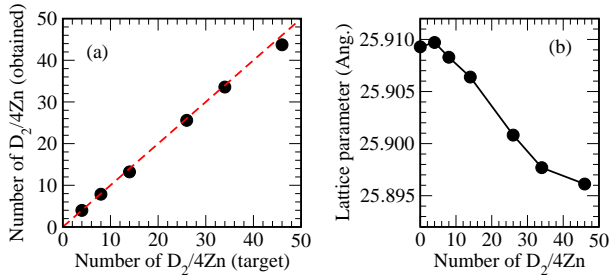} 
\caption{
(a)The target versus refined values for the hydrogen loading of MOF5
host lattice. 
(b) The lattice parameter as a function of hydrogen loading.}
\label{fig2}
\end{figure}

Due to the large incoherent cross section of H$_{2}$,
the neutron diffraction data were collected on a
deuterated-MOF5 sample, which  was 
synthesized as described in detail in Ref.\onlinecite{details}.
Because of the large cubic crystallite sizes obtained from the synthesis,
we ground the sample in a helium-filled glovebox
 prior to neutron powder diffraction to eliminate the
effects of preferred orientation.   Figure~1 shows the diffraction data from 
the deuterated-MOF5 host lattice which was obtained on BT-1 at NIST. The pattern was
taken over a period of 8 h with the 2.5 gram sample loaded into a vanadium cell
at 3.5 K. The agreement between the data and refinement is
excellent. We note that the neutron diffraction data allowed the determination
of the hydrogen atom positions of the BDC linker, which were not previously 
observed.

Having characterized the MOF5 host lattice,
we next studied the adsorption of hydrogen in MOF5 as a
function of D$_{2}$ concentration per formula unit (i.e. 4Zn)\cite{uf}. 
The hydrogen loading was achieved by first filling a well-known
dosing-volume to a target pressure and then
exposing it to the MOF5 sample at 70K. The sample was
then  cooled  down to 30K at which point the pressure
decreased to a negligible value as the D$_{2}$ was adsorbed. Once the system
was equilibrated at 30~K, the sample was further cooled down to 3.5 K
before the measurements. The sample was loaded with the
following concentrations;
nD$_{2}$=4,8,14, 26, 34, and 46 D$_{2}$ per molecular
formula (i.e. per 4Zn)\cite{uf}. We note that one D$_{2}$/4Zn
corresponds to about 0.255-wt\% hydrogen uptake. 
For the nD$_{2}$=46, the final pressure at 30~K was
non-zero, and we briefly pumped the system to remove 
free D$_{2}$. None of the structural
refinements of the deuterium loaded-samples showed
any evidence for solid D$_{2}$, indicating that the deuterium was
adsorbed onto the MOF5. This was further supported
by the total amount of hydrogen obtained from the
refinements as shown in Fig.~2(a). Apart from the
last point where we had to remove unadsorbed deuterium gas,
the target and refined values for the total amount of
deuterium molecules adsorbed in MOF5 lattice are in
very good agreement. It is
quite interesting to note that at cryogenic temperatures,
the MOF5 host lattice actually has enough space
to hold up to 10-wt\% hydrogen.  If we can find a way to engineer 
MOF5-H$_{2}$ interactions in these materials to
hold the hydrogen molecules in the structure at
ambient temperatures, one could easily store
enough hydrogen for practical applications.
Figure~2(b) shows that there is a small contraction
of the lattice upon hydrogen loading. We attribute
this to a small charge transfer ($\approx 0.1 e$)
to the hydrogen molecules which results in an attractive
Madelung Coulomb energy.  In the rest of this letter, we discuss where and
how the hydrogen molecules are packed into the MOF5 structure
as a function of D$_2$-loading.

\begin{figure}
\includegraphics[width=80mm]{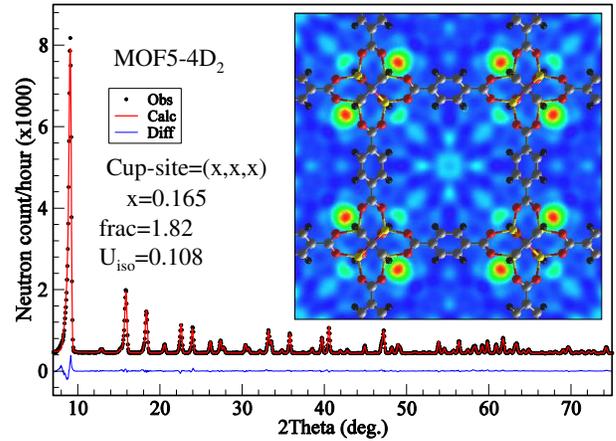}
\caption{
The neutron powder diffraction pattern ($\lambda=2.08$\AA) of the
MOF5-4D$_{2}$ at 3.5 K (dots) plus the Rietveld refinement (solid)
with space group Fm${\bar 3}$m and a=25.9097 \AA, and difference plot (bottom).
The inset shows the real-space Fourier-difference scattering-length
density superimposed with the MOF5 structure,
indicating the location of cup-sites for the
first hydrogen adsorption (red-yellow-green region).
 The refinement was characterized by $\chi^2$=2.736 and R$_{wp}$=5.54\%.}
\label{fig3}
\end{figure}

Figure~3 shows the diffraction pattern from MOF5 which was
loaded with 4D$_{2}$/4Zn. In order to locate the hydrogen
adsorption sites, we first performed a Rietveld structural refinement using
the model for the the MOF5 host structure, ignoring the adsorbed D$_2$ molecules.
The difference-Fourier scattering-length
density based upon this model, shown in the inset to Fig.~3, was then used to locate the 
adsorbed D$_2$ molecules.  The Fourier plot clearly shows where the hydrogen molecules
are adsorbed (red-yellow-green region).  Isosurfaces of the three-dimensional difference-Fourier
scattering-length density for a loading of 8D$_2$/4Zn are shown in Fig.~4(a). 
The first adsorption sites (blue) are the positions 
at the center of the three ZnO$_{3}$ triangular faces,
which resemble a cup shape and were termed the "Cup-site".  There are four such sites,
forming a tetrahedral cluster (blue region in Fig.~4a). Having
determined the location of the first-adsorption sites, 
we then further refined the structural model, 
explicitly including the D$_{2}$ molecules 
at the first adsorption site.  The positions, 
isotropic-thermal factors, and fractional occupancies 
of the adsorbed D$_{2}$ molecules were refined. 
 The deuterium molecules were treated as point
scatterers with double occupancy. The final 
refinement for a deuterium loading of 4D$_{2}$/4Zn is 
shown in Fig.~3. The 
agreement between data and the fit is very good.  
For the 4D$_{2}$/4Zn loading, we also observed a small
amount of D$_2$ (i.e. 10\% ) adsorbed at a secondary adsorption  
sites (green isosurface in Fig.~4(a)). 
For 8D$_2$/4Zn loading, these
two adsorption sites are almost fully occupied\cite{details}.
Unlike the cup-sites, the second
adsorption sites are on top of a single ZnO$_{3}$ triangles and 
were hence denoted as the "ZnO$_{3}$-site". These sites
also form a tetrahedron about the metal-oxide cluster. 
We observed that 
with further hydrogen loading (i.e. 14 and 26 D$_2$/4Zn), 
there are two additional adsorption sites
which start to populate in almost equal proportion. These sites are shown
in Fig.~4(b) as light-blue and brown spheres. The adsorption
sites just above the two oxygen ions are called the "ZnO$_{2}$-site" (see Fig.~4(b)). 
The fourth adsorption site is basically
the top of the hexagonal linkers, which we termed the "Hex-site". 
The refined fractional positions of these four sites are
summarized in Fig.~4. 
At 26 D$_2$/4Zn loading, structural refinement indicates that
these four adsorption sites are almost  totally occupied\cite{details},
yielding 6.63-wt\%  deuterium uptake.

\begin{figure}
\includegraphics[width=70mm]{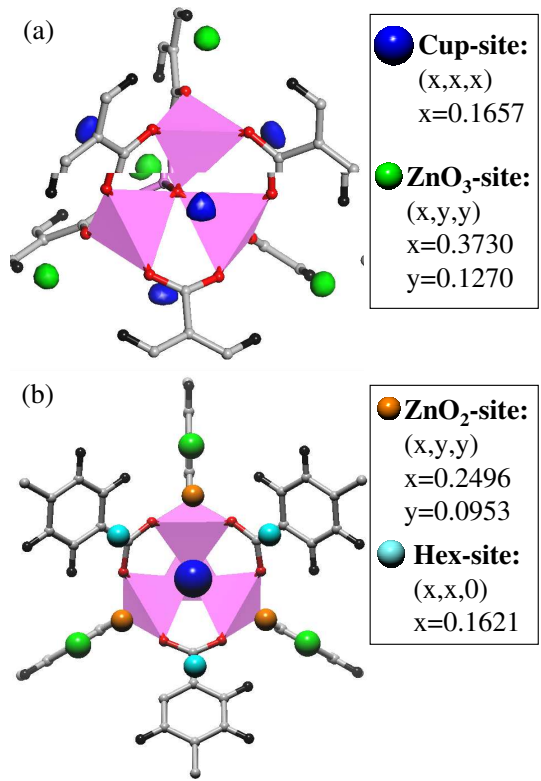}
\caption{
The hydrogen absorption sites obtained from difference
Fourier analysis.Top: The first (blue) and second (green) absorption
sites, respectively. Bottom: A view along the 3-fold axis, 
showing the four absorption sites together. In addition to the first two
absorption sites (blue and green), 
ZnO$_2$ (brown) and Hex-sites (light-blue)
are shown.}
\label{fig4}
\end{figure}

It is important to know if the adsorption sites
reported above make sense in terms of 
hydrogen host-lattice interactions and energetics.
Hence, we  have also performed  total energy
calculations from density functional theory (DFT).
The calculations were performed within the plane-wave 
implementation\cite{castep} of the local-density approximation 
(LDA) to DFT. We used Vanderbilt ultra-soft pseudopotentials\cite{usp}. 
We relax only the hydrogen molecules inside the 
primitive cell of the MOF5 structure, which contains 106 atoms\cite{uf}. 
A cutoff energy of 340 eV was found to be enough for total 
energies to converge within 0.5 meV/atom.  

\begin{table}
\begin{tabular}{|r||c|c|} \hline\hline
Cup-site & $\Delta E_{par} = 0.133$ eV & $\Delta E_{perp}=0.160$ eV \\
ZnO$_{3}$-site & $\Delta E_{par} = 0.086$ eV & $\Delta E_{perp}=0.115$ eV \\
ZnO$_{2}$-site & $\Delta E_{par} = 0.056$ eV & $\Delta E_{perp}=0.108$ eV \\
Hex-site & $\Delta E_{par} = 0.092$ eV & $\Delta E_{perp}=0.106$ eV \\ \hline\hline
\end{tabular}
\caption{The calculated binding energies for four-adsorption sites when
H$_{2}$ molecule is parallel and perpendicular to the 3-fold axis near
the adsorption sites.}
\end{table}

The energies of the four adsorption
sites are summarized in Table~1 for two different
orientation of the hydrogen molecule. 
The binding energies are in good agreement with the
experimental finding the "Cup-site" is the most 
energetically stable, followed by the "ZnO$_{3}$-site".
The calculated binding energies for the "Hex-site" and "ZnO$_{2}$-site"
are quite close each other, in agreement with the equal
population of these sites observed experimentally. Finally, we also
point out that the "Hex-site" and "ZnO$_{2}$-site" are further
stabilized by the intermolecular interactions amongst the adsorbed deuterium molecules.
 For example,
each "ZnO$_{3}$-site" is surrounded by three "ZnO$_{2}$-site" due
to local 3-fold symmetry. Upon initial hydrogen loading, the molecules
first go to the biggest cavities, namely the "cup-site"
and the corner sites (i.e. "ZnO$_{3}$-site"). After that, each hydrogen
molecules is surrounded by three hydrogen molecules
(of the "ZnO$_2$-site" and "Hex-site"). In this way, the packing
of the hydrogen molecules are optimized due to both 
H$_2$-ZnO-cluster interaction and the H$_2$-H$_2$ interactions.
Finally, we note that the adsorption energy of the first three
sites shows significant anisotropy (about 30 meV) with respect
to the orientation of the H$_2$ molecule. Therefore we expect
significant splitting of the ortho-para transitions for H$_2$ molecules
adsorbed at these three sites with a more isotropic transition for the hydrogen
molecule at the "Hex-site". These results seem to be
consistent with the available inelastic neutron data on
H$_2$ in MOF5\cite{MOF_neutron,mof5_mike}. We also observe that
the exact location  of the H$_2$ molecule at the adsorption sites 
depends  on the H$_2$ orientation within 0.2 \AA\;, 
suggesting that a proper treatment of translation-rotation 
coupling of the H$_2$ quantum dynamics is
required to explain the inelastic neutron data\cite{mof5_mike}.

\begin{figure}[t]
\includegraphics[width=80mm]{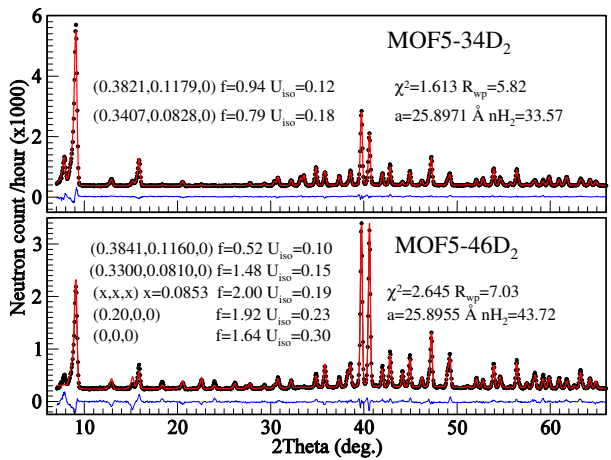}
\caption{
The neutron powder diffraction patterns (dots), Rietveld 
refinements (solid) and the difference plots (noisy line) for
high-concentration hydrogen loading, n=34 and n=46 
D$_2$/4Zn, respectively. In addition to the four adsorption sites
shown in Fig.~4,
additional hydrogen sites, occupancies and thermal factors
are also given. }
\label{fig5}
\end{figure}

So far we have discussed hydrogen loading up to
26 D$_2$/4Zn, at which point the 
adsorption sites discussed above are almost fully
occupied.   Figure~5 shows the
neutron powder diffraction patterns and Rietveld structural refinements at 
two more deuterium loadings of 34 and 46 D$_2$/4Zn, indicating
that MOF5 structure is capable of adsorbing more
hydrogens. The difference Fourier analysis indicated
that at these  high-coverages, hydrogen molecules 
form quite interesting nano-cages in the cubic-cavities
of the MOF5 structure as shown in Fig.~6. 
The first two hydrogen positions listed in Fig.~5 generate
the cubic and almost spherical D$_2$ nano-cages
shown in Fig.~6(a) and (b), respectively. 
Due to the hydrogen atoms on the organic linker, the
cubic nanocage shown in Fig.~6(a) is slightly bent along the edges.
  For 34 D$_2$/4Zn loading, these two cage
structures have about equal population; indicating some
disorder. However, with increasing the hydrogen loading
to 46 D$_2$/4Zn, we observed
that the cubic cage is destabilized with respect to the more symmetric
and exotic looking cage shown in Fig.~6(b).  
The  intermolecular distances in these nano-cages are 
on the order of 3.0 \AA, much shorter than those
found in solid H$_2$\cite{solidh2,H2_pressure}. 
At the maximum coverage of 46D$_2$/4Zn, we also determined
three additional hydrogen sites which are listed
in Fig.~5. These hydrogens  basically sit on the top
of the square faces of the nano-cage shown in Fig.~6(b),
creating quite remarkable 3D interlinked nano-cage structures.
These results suggest that the MOF host lattices may be used as a template to build
new artificial hydrogen nano-structures, which could have
quite interesting properties due to the quantum nature of
the molecules, confinement effects,
and small intermolecular distances. 
The structure of solid hydrogen under
very high-pressures has been a focus of intense research
for a long time due to theoretical predictions for 
metallic behavior\cite{H2_metal}.
Hence, we hope that our initial results for the
high-concentration hydrogen loading will give a different
perspective and direction to this important field of research\cite{H2_pressure,H2_metal}.
We  note that due to the construction of the vanadium sample holder used
in this study, we were not able to apply high pressures 
to load the MOF5 sample
with even more hydrogen molecules. The fractional occupancies
listed in Fig.~5 indicates that it should be possible
to insert more hydrogen molecules into the structure.  
We hope to perform such high-pressure studies in the near future.

\begin{figure}
\includegraphics[scale=0.45]{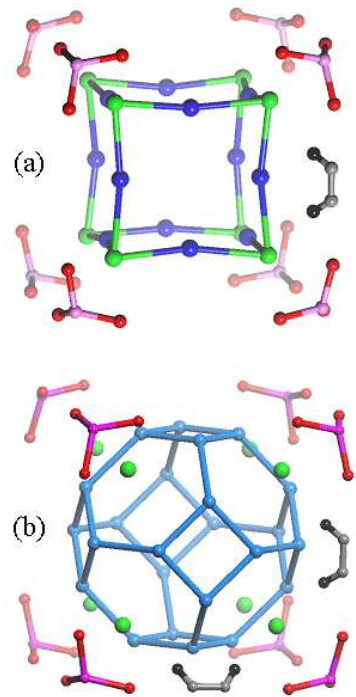}
\caption{
Two different hydrogen nano-cages  (blue and green
spheres) obtained from
Rietveld refinements at 34 and 46 D$_2$/4Zn loading,
respectively.  For clarity, only
some of the ZnO$_3$(red-pink) and CH bonds (grey-black)
 are also shown.}
\label{fig6}
\end{figure}

In conclusion, using Rietveld structural refinement 
of neutron powder diffraction data in conjunction
with difference-Fourier analysis and first-principles
calculations, we have determined the hydrogen adsorption 
sites and binding energies 
in MOF5. We have also discovered that hydrogen molecules 
form unique 3D interlinked nano-cages 
at high-concentrations of hydrogen loading. 
 Surprisingly, we find that MOF host lattice has 
enough space to hold hydrogen molecules up to 10-wt\%
at low temperatures. This implies that by using different 
organic linkers, which make hydrogen desorption difficult
 by narrowing the channels connecting the network of 
 nano-pores in MOF, one may be able to engineer these 
 materials for practical hydrogen storage at ambient conditions. 
  These results not only hold the key to optimizing MOF materials
   for hydrogen storage applications but also suggest
that MOFs can  be used as templates to create artificial 
interlinked hydrogen nano-cages with novel properties.

{\bf Acknowledgments:} 
We thank B. H. Toby and  Q. Huang for assistance with  BT-1 and
J. Eckert, J. Rowsell, and D. A. Neumann
for useful discussions.
This work was partially supported by DOE within the Center of 
Excellence on Carbon-based Hydrogen Storage Materials.


\begin{thebibliography}{99}
\bibitem{uf}
The unit formula for MOF5 is Zn$_4$O$_{13}$-(C$_8$H$_4$)$_3$.
The conventional unit cell (Fm${\bar 3}$m)
has eight of these formulas.
\end{thebibliography}
\end{document}